\documentclass[12pt,journal]{article}
\usepackage{graphicx,amssymb,amsmath,amsthm,cite,cases}
\usepackage{multicol}
\usepackage{color,multirow}
\usepackage{mathrsfs}

\def\ov{\overline}
\def\noi{\noindent}

\def\be{\begin{equation}}
\def\ee{\end{equation}}
\def\ben{\begin{eqnarray}}
\def\een{\end{eqnarray}}
\def\Da{\mathcal{D}_{\mathrm{a}}}
\def\Db{\mathcal{D}_{\mathrm{b}}}
\def\D{\mathcal{D}}

\def\R{\mathbb{R}}
\def\N{\mathbb{N}}
\def\V{\mathtt{V}}
\def\A{\mathtt{A}}
\def\B{\mathtt{B}}

\def\Z{\mathbb{Z}}
\def\til{\tilde}

\def\TP{\mathrm{TP}}
\def\FP{\mathrm{FP}}
\def\FN{\mathrm{FN}}
\def\SE{\mathrm{SE}}
\def\SEN{\mathrm{SE}_{\NN}}
\def\SEV{\mathrm{SE}_{\VV}}
\def\PP{\mathrm{PP}}
\def\PPN{\mathrm{PP}_{\NN}}
\def\PPV{\mathrm{PP}_{\VV}}
\def\AC{\mathrm{AC}}

\def\tol{\mathrm{tol}}
\def\std{{\mathrm{std}}}

\def\prdn{{\mathrm{prdn}}}

\def\VV{{\tt{V}}}
\def\NN{{\tt{N}}}

\def\vG{\mathbf{G}}

\def\vb{\boldsymbol{\beta}}
\def\vR{\mathbf{R}}
\def\vd{\mathbf{d}}
\def\vD{\mathbf{D}}

\def\vDn{\mathbf{D}_n}
\def\vDv{\mathbf{D}_v}
\def\vc{\mathbf{c}}
\def\vf{\mathbf{f}}
\def\vF{\mathbf{F}}
\def\vFt{\til{\mathbf{F}}}

\def\vg{\mathbf{g}}

\def\vC{\mathbf{C}}

\def\vW{\mathbf{w}}

\def\vRe{\mathbf{r}}
\def\vR{\mathbf{r}}
\def\vr{\mathbf{r}}

\def\vWt{\til{\mathbf{w}}}
\def\vd{\mathbf{d}}

\def\op{\hat{P}}
\def\Nq{N_q}

\def\vfq{\mathbf{f}_q}
\def\vfqv{\ov{\mathbf{f}}_q}

\newcommand{\la}{\left \langle}
\newcommand{\ra}{\right \rangle}

\DeclareMathOperator*{\argmin}{arg\,min}
\DeclareMathOperator*{\Max}{MaxIt}

\newtheorem{remark}{Remark}
\newtheorem{keywords}{\em{Keywords}}

\newcommand{\Spann}{{\mbox{\rm{span}}}}

\title{Sparsity based morphological identification of heartbeats}
\author{Laura Rebollo-Neira\\
email: l.rebollo-neira@aston.ac.uk\\
Mathematics Department\\
Aston University\\
B4 7ET Birmingham, UK\\\\
Khalil Battikh and  Amadou Sidi  Watt \\
ENSIIE - École nationale supérieure d'informatique pour l'industrie et l'entreprise\\
1 Rue de la R\'{e}sistance, 
91000 \'{E}vry-Courcouronnes, France}
\date{}
\begin{document}
\maketitle
\begin{abstract}
{\em{Background:}} The electrocardiogram (ECG) is one of the 
most common primary tests to evaluate the health of the heart.
Reliable automatic interpretation of ECG records is 
crucial to the goal of improving public health.  
It can enable a safe inexpensive monitoring.  This work 
presents a new methodology for morphological 
identification of 
   heartbeats, which is placed outside the usual 
	machine learning framework. 

{\em{Method:}} The proposal considers the sparsity of the 
representation of a heartbeat as a parameter for 
morphological identification.
 The approach involves 
greedy algorithms for selecting elements 
from redundant 
dictionaries, which should be previously learnt from 
examples of the classes to be identified. 
Using different metrics of sparsity, 
the dictionary rendering the smallest sparsity 
value, for the equivalent approximation quality of a new heartbeat, classifies the morphology of that beat. 
This study focuses on a procedure of learning  
 the dictionaries for representing heartbeats and compares 
several metrics of sparsity for morphological
         identification on the basis of those metrics.

{\em{Results:}} The suitability of the method is illustrated
by binary differentiation of Normal and Ventricular heartbeats in
 the MIT-BIH Arrhythmia  data set. In general classification 99.7\% of the 
	Normal beats and 97.6\% of the Ventricular beats in  
	the testing sets are correctly identified. In interpatient  
	assessment 91.8\% of the Normal beats and 91.0\% of 
	Ventricular beats are correctly identified. 
 Even more important than these scores
 is the fact that they are produced on the bases of 
	 a single parameter.

{\em{Conclusions}}:
The numerical tests, designed to emphasise the 
 interpretability and reliability of the approach, 
demonstrate the potential of the method to contribute 
towards the development of a well grounded expert system for 
classification of heartbeats in ECG records.

\end{abstract}
\newpage
\section{Introduction}
Cardiovascular diseases (CDVs) are a group of disorders of the heart and blood vessels,  including coronary heart disease, cardiovascular disease, rheumatic heart disease and other conditions.
These diseases are the leading cause of global death. The
World Health Organisation (WHO) estimates that 32\% of all deaths worldwide are caused by CDVs. In addition to preventing CDVs by
maintaining a healthy lifestyle, early detection can curb mortality rates for heart disease patients and also for people with increased risk of this condition.

An electrocardiogram (ECG) is a safe and  non-invasive  procedure
  that  detects cardiac abnormalities by measuring the electrical activity generated by the heart as it contracts.
Sensors attached to the skin are used to detect the electrical signals produced by the heart each time it beats. These signals are
 analysed  by a  specialist to assess if they are unusual.
Computerised ECG analysis plays a critical
role in clinical ECG workflow. It is predicted  that in the
future full automated analysis of ECG may become more reliable
than human analysis.

Automation of ECG interpretation has been successfully
 addressed  by Artificial Intelligence (AI) methods
  which fall within the end-to-end deep learning
 framework \cite{AHO17,MR19, HRH19,RRP20}. However, due to the
over-parameterised black-box nature of this framework \cite{LBH15} it is difficult to understand how deep models make decisions.   
Besides, interpretation 
using Deep Neural Networks (DNN) has been shown to be 
 susceptible to adversarial attacks \cite{HHL20,ML22}. 
 Automatic classification of heartbeats
 has also been addressed by different techniques of
 machine learning which rely on the ability to
extract distinctive features of the beats in each class. Reviews and
summaries of these techniques can be found in
\cite{LSC16,AFA17,LMM18,MM22}.

In this work we present an alternative viewpoint for the
  specific task of
 morphological identification of heartbeats. We base the proposed
  method on the ability to model the morphology of
a heartbeat as superposition of elementary components. We work under the  assumption that there exits a set of elementary component which is
  better suited for representing heartbeats of certain morphology
 than others. Inspired by a previous  work on face recognition
 \cite{WYG08}, we consider that the sets of elementary components
are redundant. However, we differ from \cite{WYG08} in the methodology  and also
in that here the redundant sets are to be learned from  data.
 The common ground between the approaches is that both base 
 the identification criterion on the concept of sparsity.

In the signal processing field, a signal approximation
is said to be sparse if it belongs to a
subspace of much smaller dimensionality than that of the space were the signal comes from. The search for the approximation subspace is carried out using  a redundant set, called a `dictionary'. Only a few
elements of the dictionary, called `atoms', are finally used for
constructing the signal approximation, called `atomic decomposition'. 
 Sparse approximation of
heartbeats is applied in \cite{RR18} for classification using Gabor dictionaries.
The particular Gabor's atoms involved in the representation of
the beats, as well as the coefficients in the atomic decomposition,
are taken as features to feed machine learning classifiers.
Contrarily, our proposal relies on the possibility  of
learning dictionaries, which play themselves a central role in the decision
making process.  To the best of our knowledge this setup has not been considered before. We focus on the layout of the approach and
  test it for binary classification.

The paper is organised as follows. Sec.~\ref{SRUD} discuses the 
greedy algorithms to be considered, as well as  the method 
for learning the dictionaries. The sparsity metrics supporting the 
discrimination criteria are also introduced in this section. The 
 numerical tests illustrating the approach are 
placed in Sec.~\ref{numex}. 
The conclusions are drawn in 
Sec.~\ref{conclu}.
\section{Signal representation using dictionaries}
\label{SRUD}
We start this section by introducing 
some basic notation. 
Throughout the paper 
$\R$ represents the set of real numbers, and  $\N$ and 
$\Z$ the sets of natural and integer numbers respectively.
Boldface lower case letters indicate Euclidean vectors and 
boldface capital letters indicate matrices. The 
corresponding components are indicated using standard mathematical fonts
e.g., $\vd \in \R^N$ is a vector of components
$d(i)\in \R,\, i=1,\ldots,N$ and $\vG \in \R^{N\times M}$ is a matrix of components $G(i,j),\ i=1,\ldots, N, \, j=1,\ldots,M$.
The Euclidean inner product $\la \cdot, \cdot \ra$ between vectors 
in $\R^N$ is defined as $$\la \vd ,\vg \ra= \sum_{i=1}^{N}
  d(i)g(i).$$
This definition induces the 2-norm 
$\|\vg\| = \sqrt{\la \vg ,\vg \ra}$. The 1-norm of a 
vector $\vc \in \R^{N}$, is indicated as $\|\vc\|_1$ and 
calculated as $\|\vc\|_1= \sum_{n=1}^N |c(n)|$. 

The inner product between matrices in $\R^{N \times M}$
is the Frobenius inner product, which is 
defined as
$$\la \vF ,\vG \ra_F= \sum_{i=1}^{N} \sum_{j=1}^{M}
  F(i,j)G(i,j).$$
This definition induces the Frobenius norm
$\|\vG\|_F = \sqrt{\la \vG ,\vG\ra_F}.$
The transpose of a matrix $\vG$ is
indicated as $\vG^\top$.

Given a signal $\vf$ as a vector in $\R^N$,  
the $K$-term
atomic decomposition for its 
  approximation is 
 of the form
\be
\label{atoK}
\vf^K=\sum_{n=1}^K c(n)\vd_{\ell_n},
\ee
where the elements $\vd_{\ell_n}$, called atoms, are
chosen from a redundant set
$\D=\{\vd_n, \in \R^N, \|\vd_n\|=1\}_{n=1}^M$,
called dictionary. The set of all linear combination of 
elements in $\D$ is denoted as $\Spann\{\D\}$.
The problem of how to select from $\D$ the
$K$ elements $\vd_{\ell_n},\,n=1\ldots,K$ such that
$\|\vf^K - \vf\|$ is minimal is intractable (there
are $\frac{N!}{(N-K)!K!}$ possibilities).
In practical applications the problem is addressed by `tractable' methods. For the most part these methods are realised by
\begin{itemize}
\item [(a)] Expressing $\vf^K=\sum_{n=1}^M c(n)\vd_n$
 using only  $K$-nonzero coefficients minimising 
 the 1-norm $\|\vc\|_1=\sum_{n=1}^M  |c(n)|$ \cite{CDS98}.

\item [(b)] Using a greedy pursuit strategy for stepwise 
selection of the $K$ elements $\vd_{\ell_n},\,
n=1,\ldots,K$, for producing the approximation 
\eqref{atoK}.
\end{itemize}

We restrict consideration to greedy pursuit algorithms, 
because for the application we are considering these types of methods are effective and faster that those based on minimisation of the 1-norm.

\subsection{Pursuit Strategies}
In the context of signal processing the simplest pursuit strategy
 is known under the name of Matching Pursuit (MP)\cite{MZ92}. 
Depending
on their
implementation and context of application 
variations of  
the  MP 
approach can also be found under different names.
 We discuss here a refinement of MP known as  
Orthogonal Matching Pursuit (OMP) \cite{PRK93} as well as  
the stepped wise 
optimised version termed 
 Optimised Orthogonal Matching 
 Pursuit (OOMP) \cite{RNL02}.

\subsubsection {From MP to OMP}
The MP algorithm evolves by successive approximations as follows:
 Setting $k=0$ and starting with an initial approximation
$\vf^0=0$ and residual $\vr^{0} = \vf$,
the algorithm
  progresses  by sub-decomposing the $k$-th order residual
in the form
\be
\vr^{k} =
\la \vd_{\ell_{k+1}}, \vr^{k} \ra \vd_{\ell_{k+1}}
+ \vr^{k+1},
\label{tech:1}
\ee
where $\vd_{\ell_{k+1}}$ is the atom corresponding to the index selected as
\be
\label{selMP}
\ell_{k+1}=  \operatorname*{arg\,max}_{\substack{n=1,\ldots,M}} |\la \vd_{n} , \vr^{k} \ra|.
\ee
This atom is used to update the approximation $\vf^k$ as
\be
\label{upfk}
\vf^{k+1} = \vf^{k} + \la \vd_{\ell_{k+1}}, \vr^{k} \ra  \vd_{\ell_{k+1}}.
\ee
From \eqref{tech:1} it follows that
$\|\vr^{k+1}\| \le \|\vr^{k}\|$, since
\be
\|\vr^{k}\|^{2} = |\la \vd_{\ell_{k+1}}, \vr^{k} \ra|^{2} + \|\vr^{k+1}\|^{2}.
\label{tech:2}
\ee
It is easy to prove that in the limit $k \rightarrow \infty$, the sequence  $\vf^{k}$
given in \eqref{upfk} converges to $\vf$, if $\vf \in V_{M}=\Spann\{\D\}$, or to
$\op_{V_{M}}\vf$ the orthogonal projection of $\vf$ onto
$V_{M}$, if $\vf \notin V_{M}$ (a pedagogical proof can be found on \cite{RNRS20}). However, the  method is not 
 stepwise optimal because it does not
yield an orthogonal projection at each
step.  Accordingly, the algorithm may select linearly dependent atoms, which is the main drawback of MP when applied with highly coherent dictionaries. 
A refinement to MP, which does yield an
orthogonal projection approximation at each step is called 
OMP \cite{PRK93}. The method selects the atoms as in 
\eqref{selMP} but at each iteration produces a decomposition of the signal as given by:
\begin{equation}
\vf^k = \sum_{n=1}^{k} c^k(n) \vd_{\ell_{n}} + \tilde{\vRe}^{k},
\label{tech:5}
\end{equation}
where the coefficients $c^k(n)$ are computed in such a way 
that
it is true that
$$\sum_{n=1}^{k} c^k(n) \vd_{\ell_{n}}= \hat{P}_{\V_{k}}\vf,\quad{\text{with}}
\quad V_{k}= \Spann\{\vd_{\ell_{n}}\}_{n=1}^k.$$
The superscript of $c^k(n)$ in \eqref{tech:5} indicates
the dependence of these quantities on the iteration step $k$.
Thus,
in addition to selecting linearly independent atoms,
OMP yields the unique element $\vf^k \in V_{k}$ minimising
$\|\vf^k -\vf\|$. As discussed  next,
a further refinement to OMP, called 
OOMP \cite{RNL02}, selects also the atoms in 
order to minimise in 
a stepwise manner the norm of the residual error.
\subsubsection{The OOMP method}
The OOMP approach iterates as follows. 
The algorithm is initialised by setting:
$\vr^0=\vf$, $ \vf^0=0$, $\Gamma= \emptyset$
and $k=0$. The
 first atom
 is selected as the one corresponding to the index
$\ell_{1}$ such that
\be
\ell_{1}=\operatorname*{\,max}_{n=1,\ldots,M}
 \left |\la \vd_n,\vr^{0}\ra \right|^2.
\ee
This first atom is used to
assign $\vW_1=\vd_{\ell_{1}}=\vb_1$,
calculate $\vr^{1}= \vf - \vd_{\ell_{1}} 
\la \vd_{\ell_{1}}, \vf\ra$, and  iterate
 as prescribed below.

\begin{itemize}
\item[1)] Upgrade the set $\Gamma \leftarrow  \Gamma \cup 
\ell_{k+1}$, increase $k \leftarrow k +1$, and
 select the index of a new atom for the approximation as
\be
\label{oomp}
\ell_{k+1}=\operatorname*{\,max}_{\substack{n=1,\ldots,M\\ n\notin \Gamma}}
 \frac{|\la \vd_n,\vr^{k} \ra|^2}{1 - \sum_{i=1}^{k}
|\la \vd_n ,\vWt_i\ra|^2},
 \quad \text{with} \quad \vWt_i= \frac{\vW_i}{\|\vW_i\|_2}.
\ee
\item[2)]
Compute the corresponding new vector $\vW_{k+1}$ as
\be
\begin{split}
\label{GS}
\vW_{k+1}= \vd_{\ell_{k+1}} - \sum_{n=1}^{k} \frac{\vW_n}
{\|\vW_n\|^2_2} \la \vW_n, \vd_{\ell_{k+1}}\ra.
\end{split}
\ee
including, for numerical accuracy,  the
re-orthogonalisation step:
\be
\label{RGS}
\vW_{k+1} \leftarrow \vW_{k+1}- \sum_{n=1}^{k} \frac{\vW_{n}}{\|\vW_n\|_2^2}
\la \vW_{n} , \vW_{k+1}\ra.
\ee

\item[3)]
 Upgrade vectors $\vb_n^{k}$ as                                     
\be
\vb_{k+1}^{k+1}= \frac{\vW_{k+1}}{\|\vW_{k+1}\|^2}, \quad
\vb_{n}^{k+1}= \vb_{n}^{k} - \vb_{k+1}^{k+1}\la \vd_{\ell_{k+1}}, \vb_{n}^{k}\ra,\quad n=1,\ldots,k.
\ee
\item [4)] Upgrade vector $\vr^{k}$ as
\ben
\vr^{k+1} &=& \vr^{k} - \la \vW_{k+1}, \vf \ra  \frac{\vW_{k+1}}{\|\vW_{k+1}\|^2}.
\een
\item[5)] If for a given $\rho$ value
  the condition
 $\|\vr^{k+1}\| < \rho$ has been met 
 stop the selection process. Otherwise repeat steps 1) - 5).
\end{itemize}
Calculate the coefficients  in \eqref{atoK} as
$$\vc=\la \vb,\vf\ra.$$
Calculate the final approximation $\vf^k$ as
$$\vf^k = \vf -\vR^k.$$
\begin{remark} 
The difference between  our implementations of OMP and
OOMP is only that for OMP the denominator in
\eqref{oomp} is eliminated.  The 
selection is realised as in MP (c.f.  \eqref{selMP}) 
which is not a stepwise optimal selection. Instead, 
	\eqref{oomp}  selects the atom $\vd_{\ell_{k+1}}$ which, fixing the previously selected atoms $\vd_{\ell_i},\,i=1,\ldots,k$,  
	minimises the distance $\|\vf^{k+1} -\vf\|$. The proof can be found in \cite{RNL02}.
\end{remark}
\subsection{Sparsity criterion for morphological feature 
extraction of heartbeats}
\label{SM}
\begin{figure}[h]
\centering
        \includegraphics[height=.4\textheight]{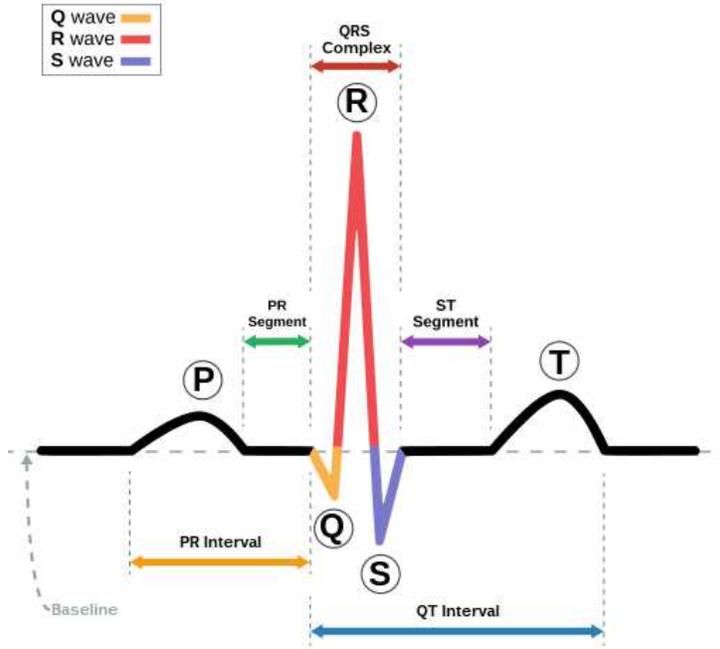}
\caption{Public domain graph of the QRS complex and P and T waves representing a normal heartbeat.}
\label{qrs}
\end{figure}

A digital ECG signal is a  
 sequence of
heartbeats, each of which is characterised by a
combination of three graphical deflections, known
as QRS complex, and the so called P and T waves. 
An idealised illustration of these deflections are 
given in Fig.~\ref{qrs}.

\begin{figure} [!h]
\centering
        \includegraphics[height=.2\textheight]{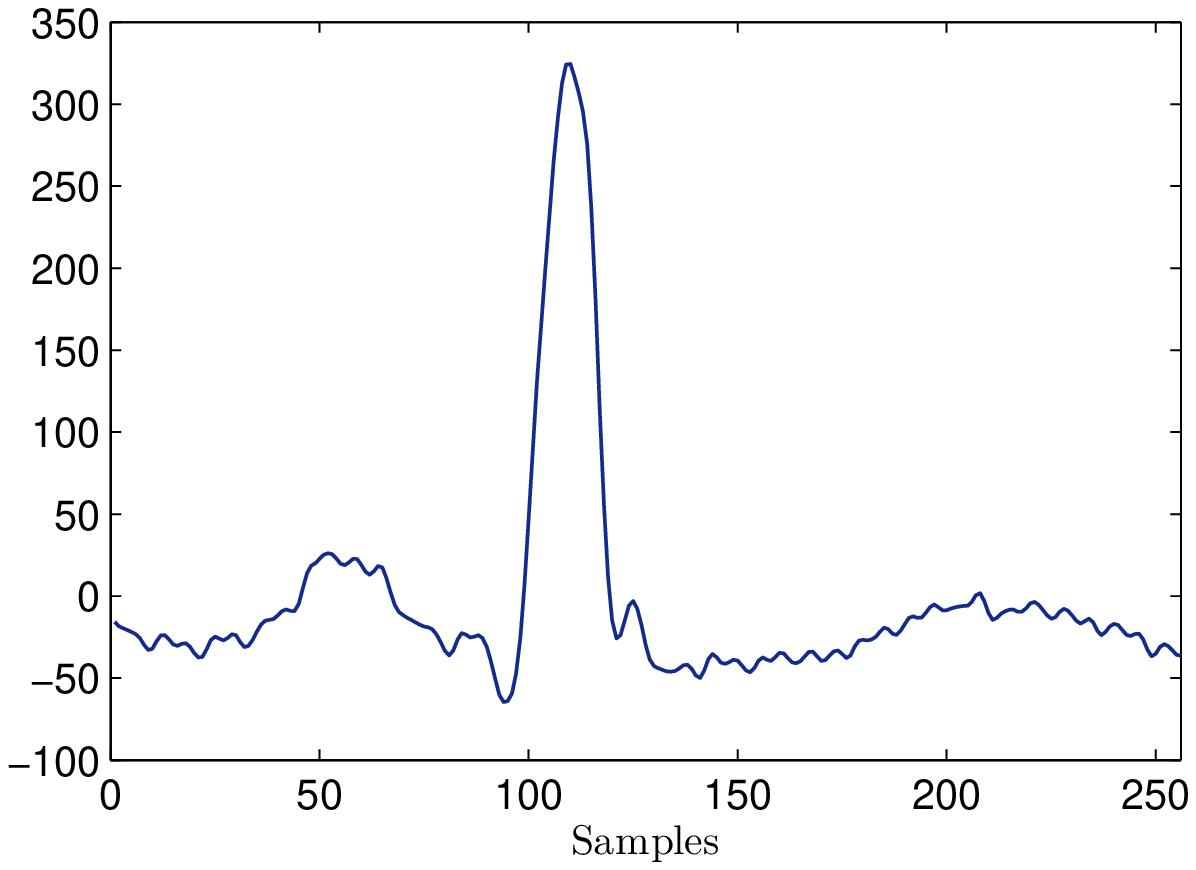}
        \includegraphics[height=.2\textheight]{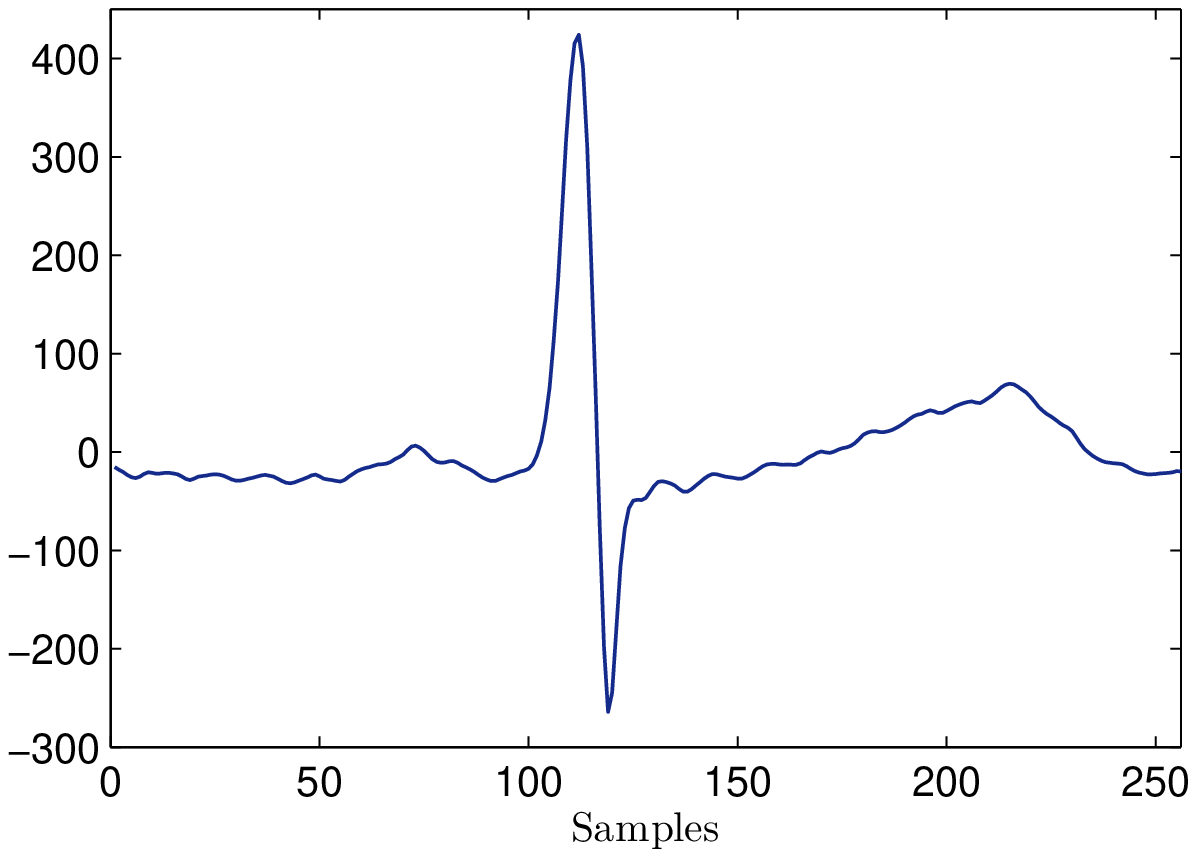}
\caption{Two normal heartbeats from different 
ECG records}
\label{beats}
\end{figure}

However, as illustrated in Fig.~\ref{beats}, in real ECG 
records the shape of the beats in the same class may vary. 

Approximation for morphological feature extraction of a heartbeat 
requires the segmentation of the QRS complex. 
In this work the segmentation is realised 
by taking  a fixed number of samples to left and right of
the location of the  R-peak, to have beats 
of equal support, say $\Nq$. 

Assuming that one wishes to morphologically differentiate heartbeats of, say class $\A$, from heartbeats of class $\B$, and that 
a dictionary $\Da$, specially
constructed to approximate a heartbeat of
class $\A$, and a dictionary $\Db$,  specially  
constructed to approximate a heartbeat
of class  $\B$, are given, we discuss next  
several possibilities of using sparsity as a distinguishability 
criterion.

In order to decide whether a heartbeat 
 $\vf \in \R^{\Nq}$ belongs to class $\A$ or class $\B$ this 
beast is approximated, up to the same precision, using 
both dictionaries, i.e. 
\be
\label{K1}
\vf^{a}=\sum_{n=1}^{K_a} c^a(n)\vd_{\ell_n^a},
\ee
where the atoms 
 $\vd_{\ell_n^a}$ are
chosen from $\Da$ and 
\be
\label{K2}
\vf^{b}=\sum_{n=1}^{K_b} c^b(n)\vd_{\ell_n^b},
\ee
where the atoms $\vd_{\ell_n^b}$ are
chosen from $\Db$. \\
\\ 
\subsubsection*{Criterion I (a) and (b) }
\label{crit}
If $K_a < K_b$ the beat $\vfq$ is assigned to class $\A$.\\
If $K_b < K_a$ the beat $\vfq$ is assigned to class $\B$.\\
If $K_a=K_b$ the criterion does not make a decision.\\

In the event that $K_a=K_b$ a decision could still be made
by recourse to a different metric.\\

\noi
(a) {\em{Smaller entropy criterion:}}
Assigning $p^a(n)=\frac{|c^a(n)|}{\|\vc^a\|_1}$ and $p^b(n)=\frac{|c^b(n)|}{\|\vc^b\|_1}$ calculate the corresponding Shannon's entropies
$$S^a= -\sum_{n=1}^{K_a} p^a(n) \ln(p^a(n)),$$
$$S^b= -\sum_{n=1}^{K_b} p^b(n) \ln(p^b(n)).$$
The entropy of, say $p^a(n),\,n=1,\ldots,K_a$, would be smaller 
if the components are fewer and the magnitude of 
some  $p^a(n)$ much larger than others.
In particular the minimum entropy value occurs 
if $p^a(n)=\frac{|c^a(n)|}{\|\vc^a\|_1}=1$ for some 
$n$ and zero otherwise, in which case $S^a=0$. Accordingly, if
$S_a < S_b$ the beat $\vf$ might be assigned to class $\A$
and if $S_b < S_a$ the beat $\vf$ might be assigned to class $\B$.
Let us recall that the criterion of smaller entropy was
 introduced for basis selection in 
the context of wavelet packets \cite{CW92}.\\

\noi
(b) {\em{Smaller norm-1 criterion.}}\\
This criterion is in line with the 
{\em{basis pursuit approach}} \cite{CDS98} 
which adopts the minimisation of the norm-1 as a way of 
producing a tractable sparse solution from a given 
dictionary.  Consequently, if
 $\|\vc^a\|_1 < \|\vc^b\|_1$ the beat $\vf$ is assigned to 
class $\A$
and if $\|\vc^b\|_1 < \|\vc^a\|_1$ the beat $\vf$ if be assigned to class $\B$.
\subsubsection*{Criterion II}
Use always the smaller entropy criterion.
\subsubsection*{Criterion III}
Use always the smaller norm-1 criterion.
\subsection{Dictionary learning}
So far we have assumed to know the dictionaries
$\Da$ and $\Db$, best suited for each class of heartbeat.  
In this section we discuss how these dictionaries can be learnt from a data set of annotated heartbeats.

The adopted  strategy to learn a dictionary, as a matrix $\vD$,  
using the heartbeats placed in an array $\vF \in \R^{\Nq \times Q}$ 
 proceed 
through a 2 step process.\\

\noi
{\em{Step 1}} 
\begin{itemize}

\item Set $k=0$. Given an initial dictionary,  as a 
matrix $\vD^k \in \R^{\Nq \times M}$, use a greedy algorithm to approximate 
 each beat  $\vfq \in \R^{\Nq}$, for  $q=1,\ldots,Q$, 
as in \eqref{K1} or \eqref{K2}  and collect the  vectors of 
 coefficients $\vc_q^k \in \R^{K_q}$.

\item Place each 
	vector $\vc_q^k$ as a column of a matrix $\vC^k \in \R^{M\times Q}$ having nonzero elements $C{({\ell_n^q},q)}^k={c_q(n)}^k,\,n=1,\ldots,{K_q}^k,\,
q=1,\ldots,Q$ and  set 
 all the other elements equal to zero. 

\item Using matrix $\vC^k$ and dictionary $\vD^k$ calculate 
the approximation $\vFt$ of $\vF$ as
$$\vFt^k =\vD^k  \vC^k$$

\end{itemize} 

\noi
{\em{Step 2}}
\begin{itemize}

\item Using the approximation $\vFt^k$ find the updated dictionary $\vD^{k+1}$ minimising 
	$\|\vFt - \vFt^k\|_F^2$. 
Thus,
\be
\label{dic}
\vD^{k+1}= 
		\argmin_{\substack{\vD}}\|{\vFt} -  {\vD} \vC^k \|_F^2 = {\vFt} {\vC^k}^\top (\vC^k{\vC^k}^\top)^{-1}.
\ee

\item Given a maximum  number of iterations,  $\Max$ say, 
and a tolerance $\tol$  for the error norm, if  $\|\vD^{k+1} - \vD^{k}\|_F < \tol$  or  $k+1 >\Max$  stop. Otherwise set $k \to k+1$ and repeat {\em{Steps 1}} and {\em{2}}.
\end{itemize} 

\begin{remark} 
The dictionary updating equation \eqref{dic}  requires
	that matrix $\vC^k{\vC^k}^\top$ should have an inverse. This would not be true if some elements in the dictionary were not chosen in the previous step. In that case, the unselected atoms 
should be 
removed from the dictionary before implementing equation 
\eqref{dic}. In practice, as long as          
$Q$ is significantly larger than $M$ and the examples are 
independent, 
	matrix $\vC^k{\vC^k}^\top$ has an inverse.
\end{remark}

\begin{remark}
 The problems of determining matrix $\vC^k$ at Step 1
and  matrix $\vD$ at Step 2 are 
convex problems with unique solution. However, the 
combined problem of determining matrix $\vC^k$ and $\vD$
 is not jointly convex. Thus, the solution depends on 
the initial dictionary. As will be demonstrated by the 
 simulations, this in not crucial within the context of 
 the proposed approach. 
\end{remark}

\section{Binary morphological differentiation of 
heartbeats}
\label{numex}
In this section we test the proposal by differentiating the 
classes $\NN$ (Normal) and $\VV$ (Ventricular Ectopic) 
in the MIT-BIH Arrhythmia data set \cite{MITDB, MM01}. 
As per the recommendations given by the Advancement of Medical Instrumentation (AAMI) the MIT-BIH Arrhythmia database is projected into five classes. Within these classes the $\NN$ beats considered here
include: The normal beats N, the
Left Bundle Branch Block Beats (LBB)   and
the Right  Bundle Branch Block  Beats (RBB).
The $\VV$ class comprises: the Premature Ventricular Contraction (PVC)  and Ventricular Escape Beats (VEB).
While the number of $\VV$ beats is much less 
than the number of $\NN$ beats, the former are 
still enough to learn the corresponding dictionary. 
The locations of the R-peaks are retrieved from the 
annotations provided with the dataset, using 
the Matlab software \cite{GAG00} available on \cite{MITDB}. The peaks are then segmented by taking $145$ samples to the right and 
110 samples to  the left of the 
 R-peak location. 
 Before learning the dictionary a quick check of the 
 training set is realised,  in order to find out 
 if there are some peaks 
 that would not qualify as 
 `good' examples. For this end we proceed as explained below. 
\subsubsection*{Screening the data training sets}
For automatic screening of a set
the segmented beats of the same class
are approximated, up to the same quality,
using a greedy algorithm and a general wavelet dictionary.
We illustrate the process by giving the details for screening 
the training set for the numerical Test I.

\begin{figure} [!ht]
\centering
        \includegraphics[height=.2\textheight]{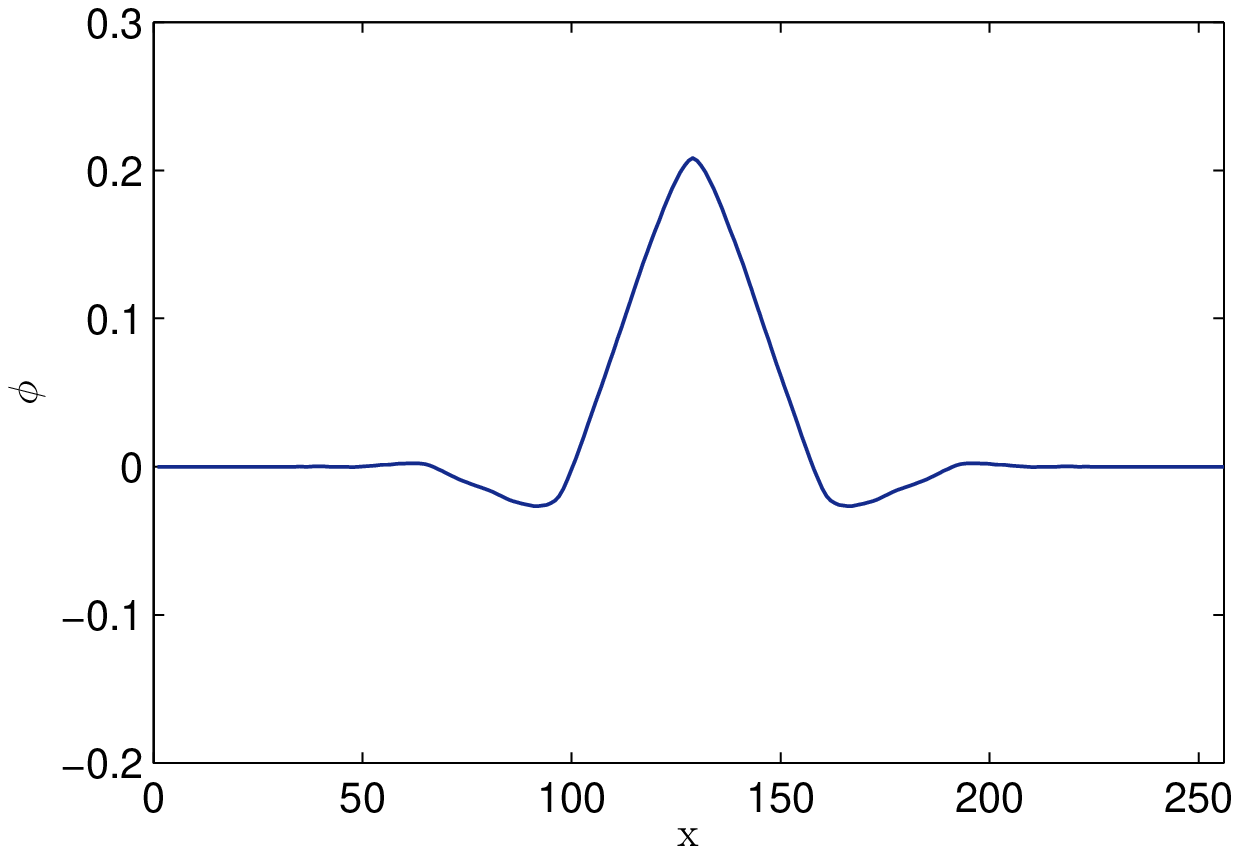}
        \includegraphics[height=.2\textheight]{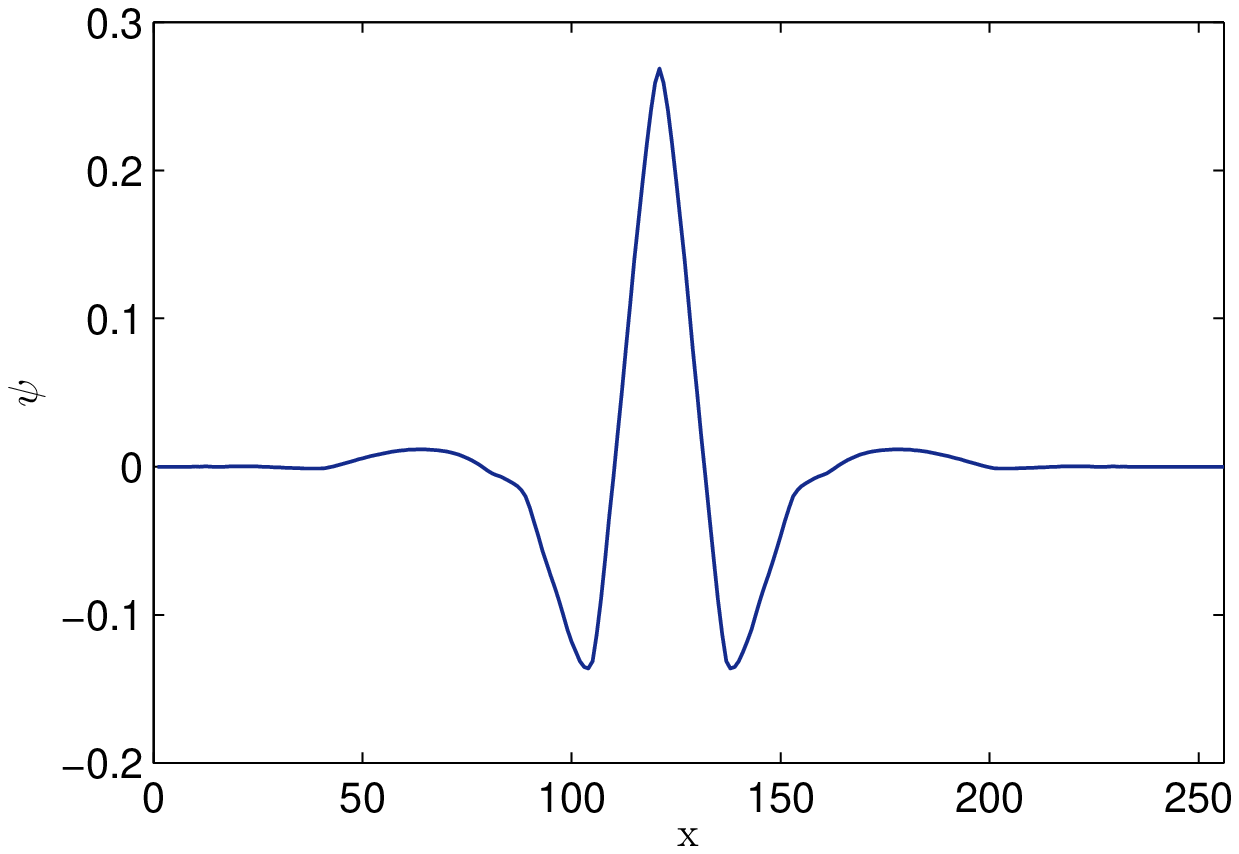}
\caption{Graph of the scaling prototype $\phi(x)$ (left)
and wavelet prototype $\psi(x)$ (right)
	for the {\bf{cbf97}} family.}
  \label{cdf}
\end{figure}

Wavelet dictionaries arise from translations,  by
a parameter $b =2^{-l},\,l \in \N$  of
scaling prototypes  \cite{ARN08,RNC19}
\begin{equation} \label{scaling_functions_dictionary}
\phi_{j_0,k,b} \left( x \right)  =  \phi  \left( 2^{j_0} x - b k \right), \quad k \in \mathbb{Z}, 
\end{equation}
and wavelet prototypes at different scales
\begin{equation} \label{wavelets_dictionary}
 \psi_{j,k,b} \left( x \right) =  \psi \left( 2^j x - b k \right), \quad
k \in \mathbb{Z}, \quad j \geq j_0.
\end{equation}
For screening of the training set we have used
the 9$/$7 Cohen-Daubechies-Feauveau ({\bf{cbf97}}) wavelet
family with $b=0.25$, $j_0=2$ and $j=3,4,5,6$,
which introduces a redundancy factor  of 2.67.
 This dictionary was
generated with the software described in \cite{CRN21} available on
\cite{webpage1}. The scaling and wavelet prototypes for
 the {\bf{cbf97}} wavelet family
 are shown in Fig.~\ref{cdf}.

\begin{figure} [!ht]
\centering
        \includegraphics[width=0.5\textwidth,height=.25\textheight]{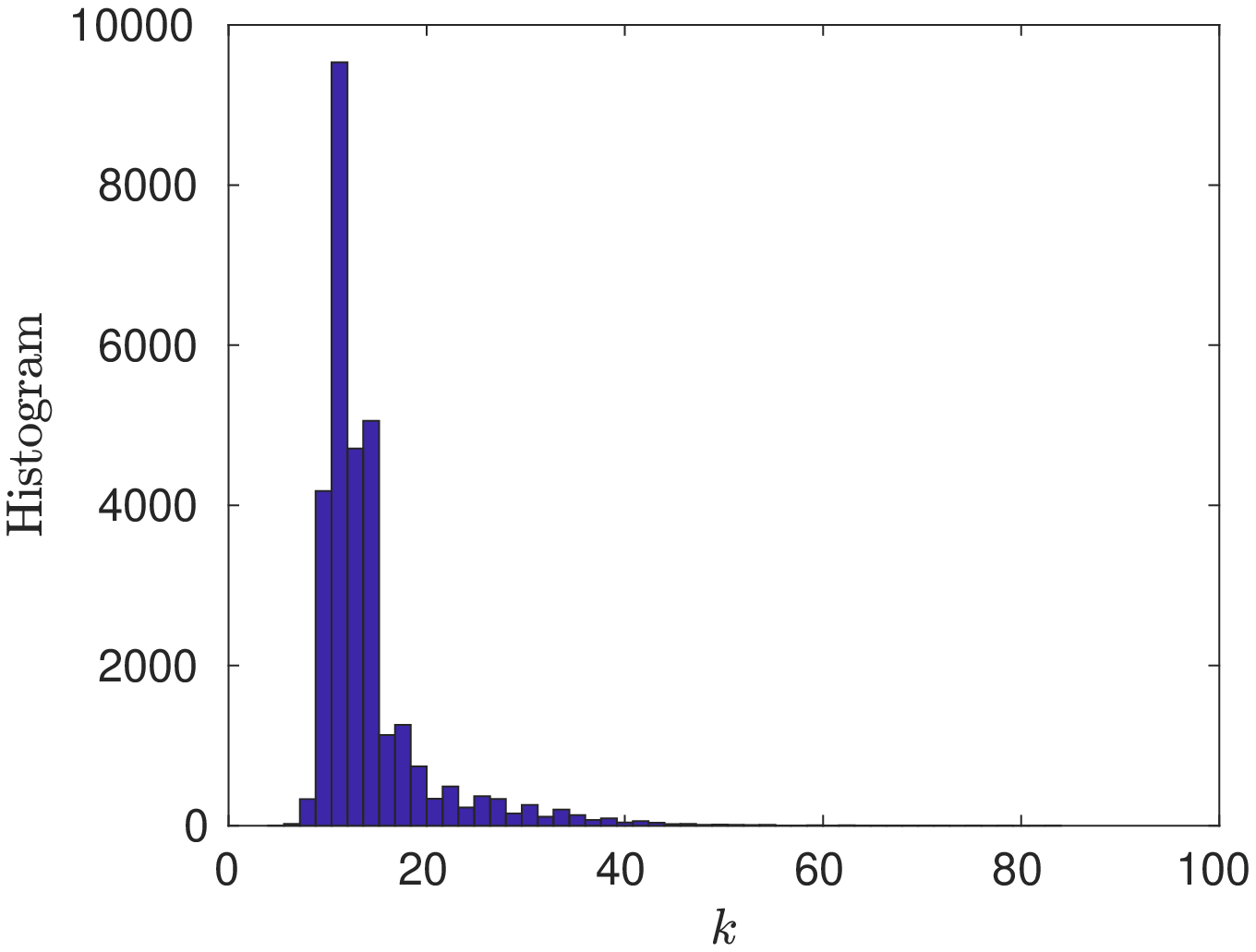}
        \includegraphics[height=.25\textheight]{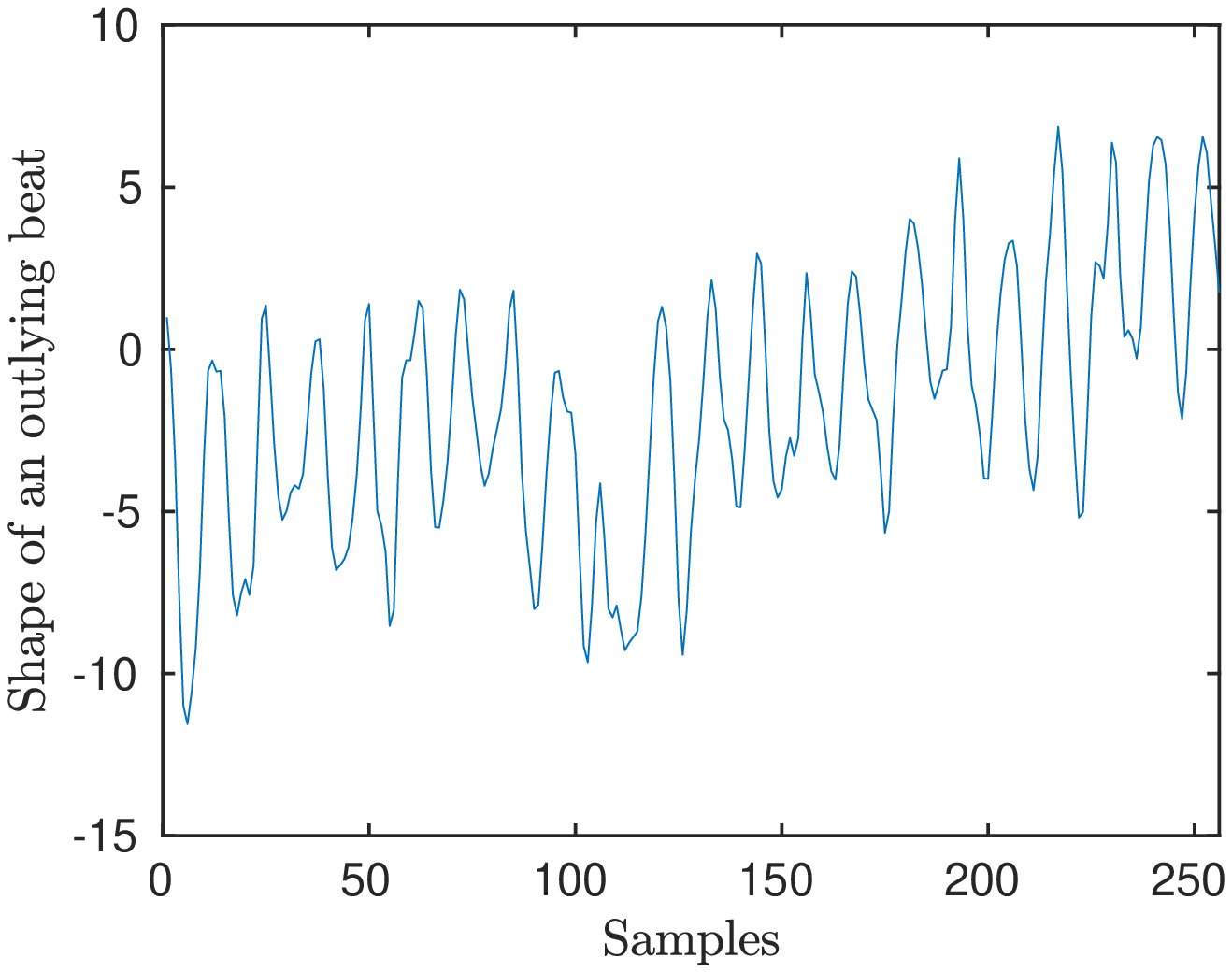}
  \caption{Histogram of $k$ values for approximating 
	(up to $\prdn=9$) 
 the $\NN$-signals 
in the training set (left graph).
Signal corresponding to the outlying value $k=63$ (right graph).}
  \label{his1}
\end{figure}

The quality in the approximation
of each beat is fixed
using the $\prdn$ metric as defined  by
$$\prdn(q)= \frac{\|\vfq - \vfq^k\|}{\|\vfq - \vfqv \|} \times 100\%, \quad q=1,\ldots,Q,$$
where each $\vfq$ is a heartbeat, $\vfq^k$ its corresponding approximation by $k$ atoms, and $\vfqv$ the mean value of
$\vfq$.
The approximation of different beasts, up to the same $\prdn$, is achieved for different values of $k$.
The left graph of Fig.\ref{his1} shows the histograms of
$k$ values obtained when approximating, up to
$\prdn=9$, the $\NN$ beats in the training set of the numerical 
Test I (c.f. Sec.~\ref{TestI}).

Even if not very noticeable in the histogram there are
a few values of $k$ very far from the main support.
These values are from `rare' signals that should be
 investigated. In this set they are just signals looking as pure
 noise, e.g. the signal in the right graph of Fig \ref{his1}
corresponds to
$k=61$, very far from the mean  value, $\ov{k}=14.31$
with standard deviation $\std=5.90$.
When learning the dictionary for
classification of $\NN$ beats we disregard beats in the
training set producing values of $k$ outside the range
 $(\ov{k} - 3\, \std, \ov{k} + 3\, \std)$.
This amounts to disregarding $2.58\%$
of the total $\NN$ beats in the training set set.
\begin{figure} [h!]
\centering
        \includegraphics[width=0.47\textwidth,height=.25\textheight]{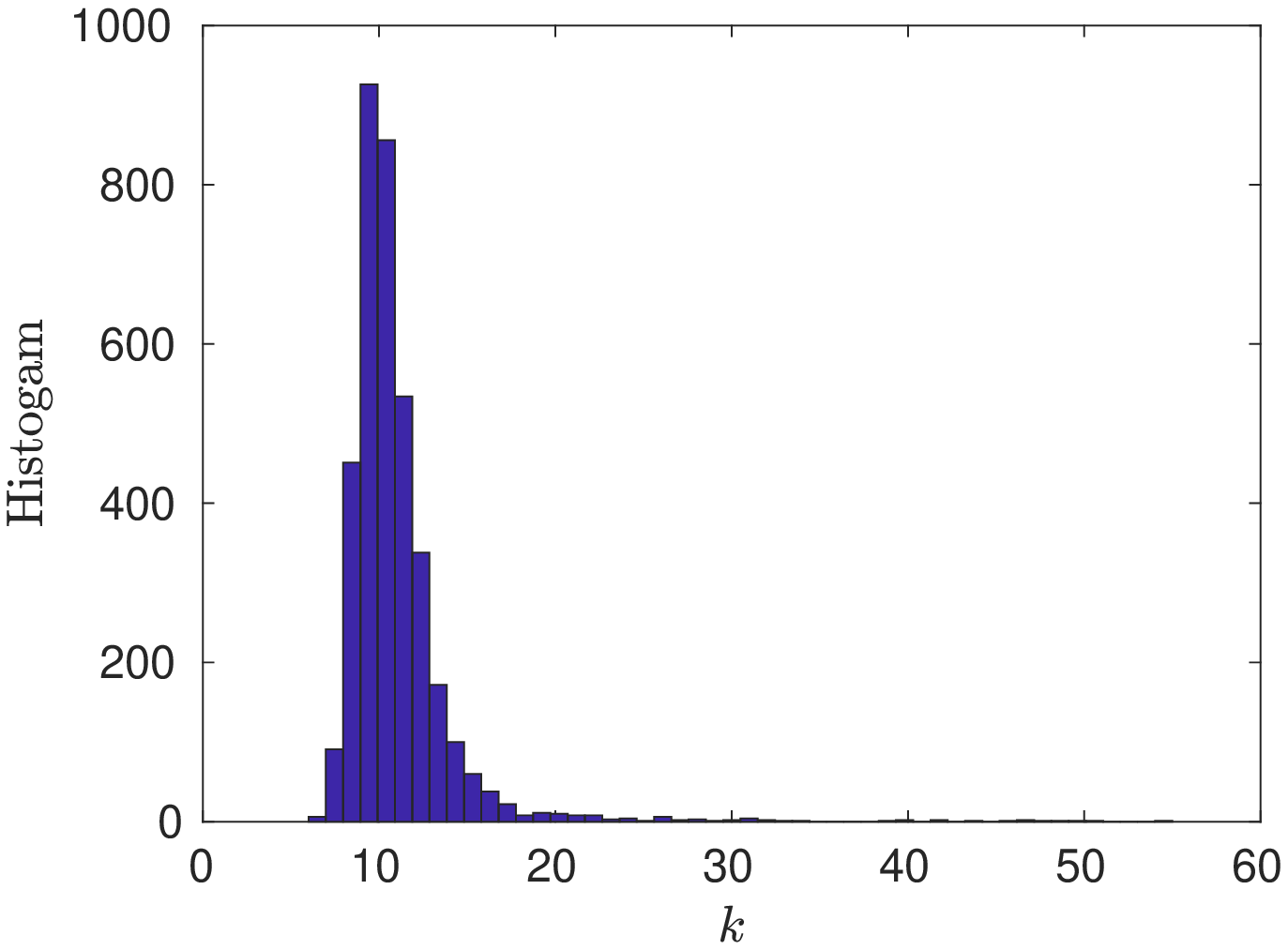}
        \includegraphics[height=.25\textheight]{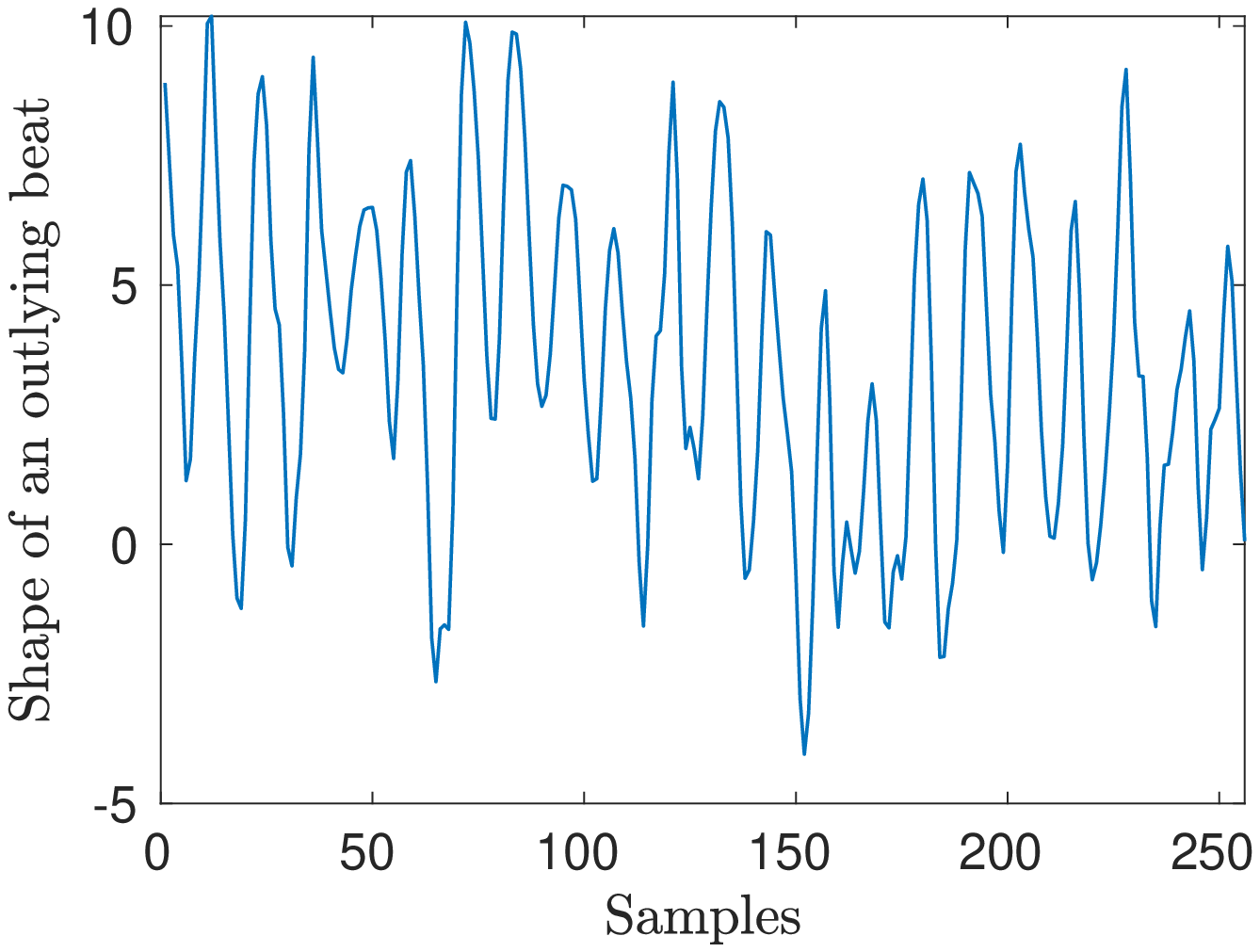}
  \caption{Histogram of $k$ values for  approximating 
	(up to $\prdn=9$) the $\VV$-signals 
in the training set (left graph).
Signal corresponding the outlying value $k=41$ (right graph).
}
  \label{his2}
\end{figure}

The left graph of Fig.\ref{his1} shows the histograms of
$k$ values obtained when approximating, up to
$\prdn=9$, the $\VV$ beats in the training set of the numerical Test I. 
For this class  $\ov{k}=10.61$ with $\std= 3.28$.
 Because the  data set of $\VV$ beats is  
 smaller than the previous one, 
when learning the dictionary for
classification of $\VV$ beats we disregard beats in the
training set outside the range
 $(\ov{k} - 2\,\std, \ov{k} + 2\,\std)$.
This amounts to disregarding $2.65\%$
of the $\VV$ beats in the training set.

\subsection{Numerical Test I}
\label{TestI}
The purpose of this test is to assess the suitability  of 
binary morphological differentiation of 
heartbeats using dictionaries learned from 
examples of $\NN$ and $\VV$ shapes 
in the dataset. For this end we randomly split the 
 $Q=89840$  $\NN$ beats into two groups: 
 35\% of the $\NN$ beats,   
are used for training the dictionary, $\vDn$ say.
The remaining 65\% of the $\NN$ beats are reserved 
for testing.
Since the $\VV$ beats are much less than the 
$\NN$ ones, $50\%$ of the $\VV$ beats are 
used for training the dictionary $\vDv$, and the remaining
 $50\%$ for testing (c.f. Table~\ref{Table1})

\begin{table}
\begin{center}
\begin{tabular}{|l|c |c || c| c||}
\hline
Sets & \multicolumn{2}{|c||}{\rm{Training}} &
\multicolumn{2}{|c||}{\rm{Testing}}\\ \hline \hline
Class  & $\NN$ & $\VV$ & $\NN$ & $\VV$ \\ \hline \hline
        Number of beats & 30000 & 3359 & 57277&  3359\\ \hline \hline
\end{tabular}
\caption{Number of heartbeats of classes $\NN$ and $\VV$
in the training and testing sets for Test I.}
\end{center}
\label{Table1}
\end{table}

The dictionaries are learned 
to have redundancy two, 
i.e. each dictionary is a matrix of real numbers of 
 size $256 \times 512$. We test the method against:\\ 
(i) The initial dictionary.\\
(ii) The different greedy algorithms considered in this 
work: MP, OMP, OOMP.\\

\begin{figure} [!ht]
\centering
        \includegraphics[height=.25\textheight]{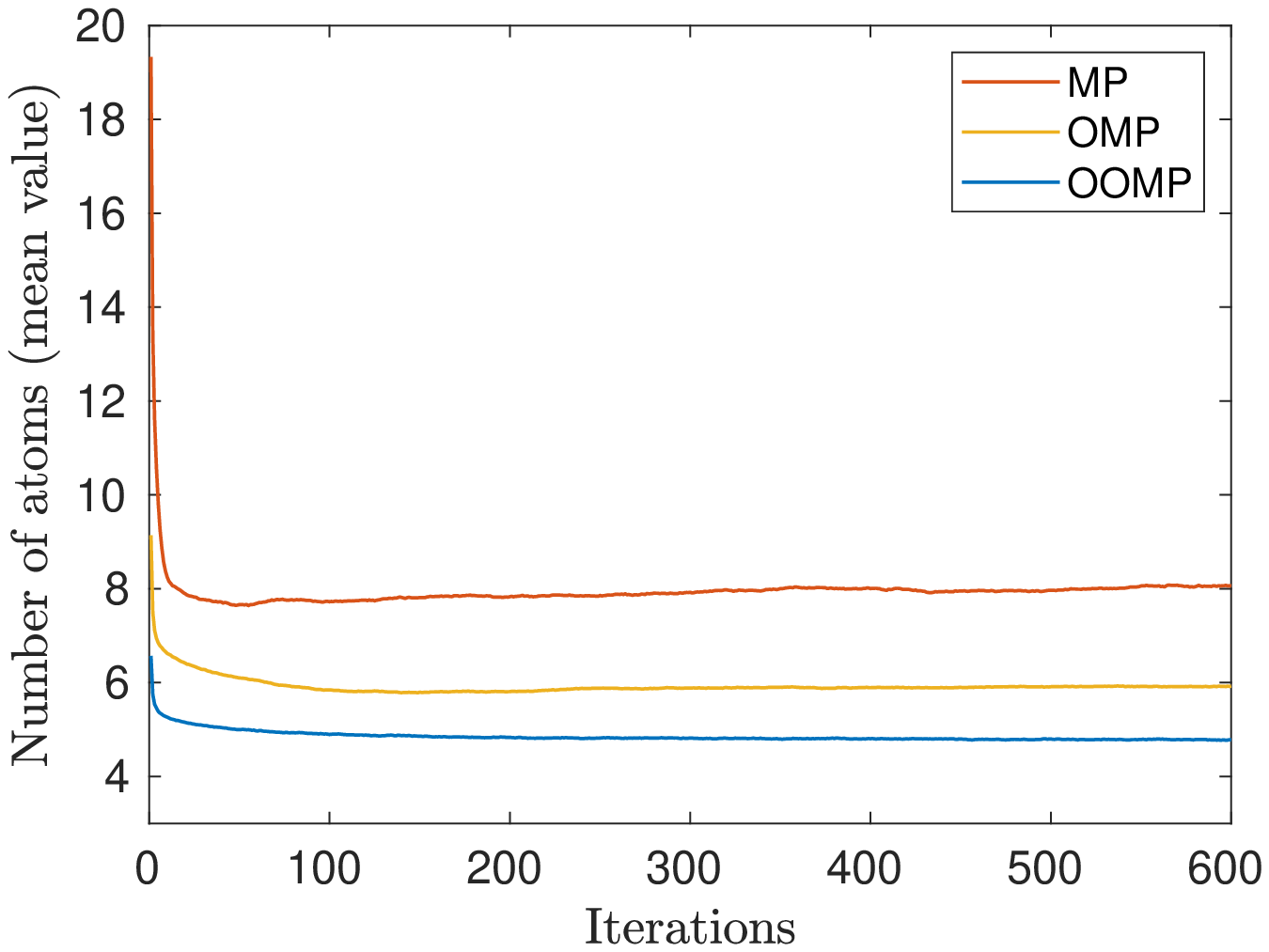}
        \includegraphics[height=.25\textheight]{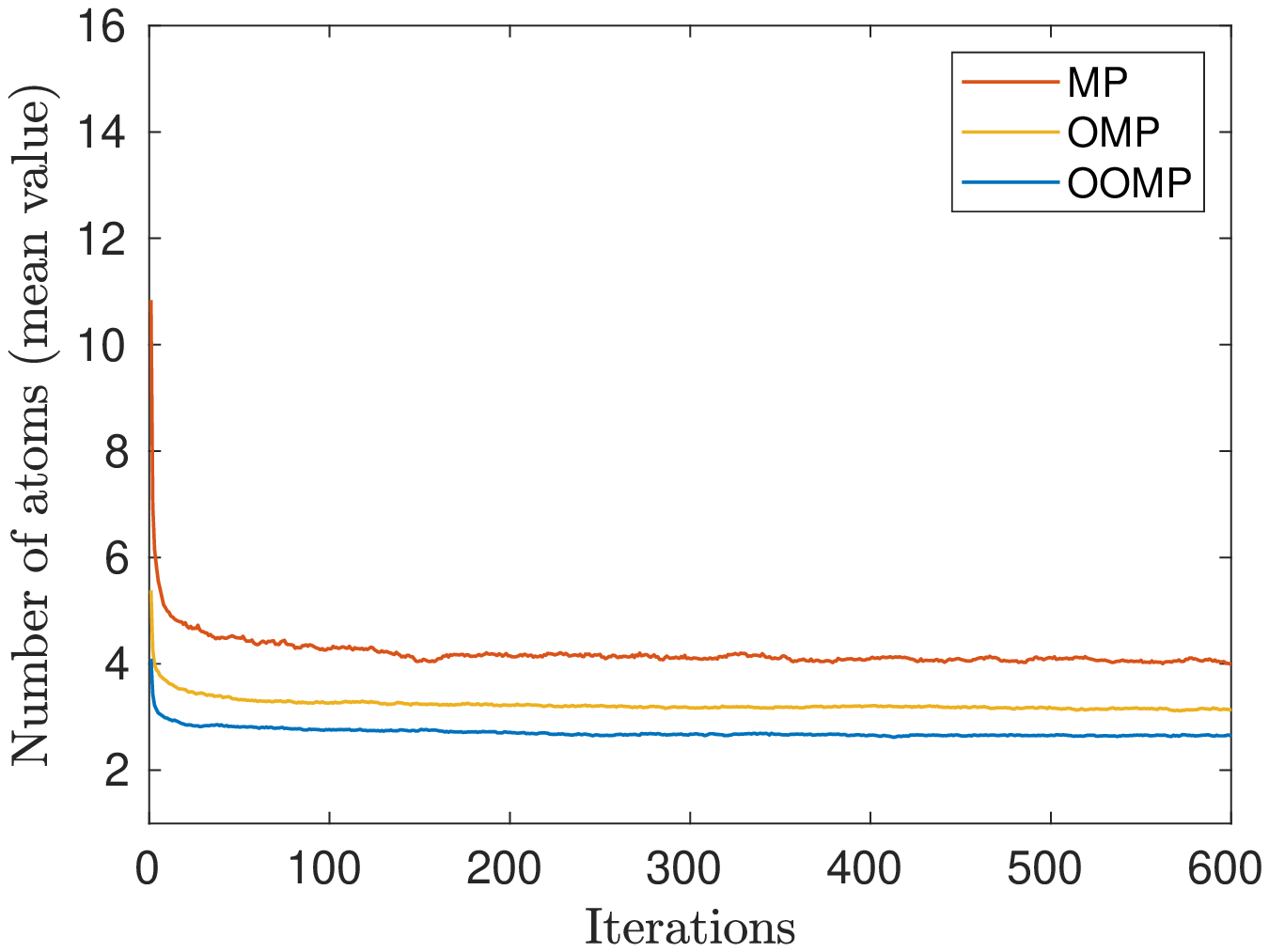}
  \caption{Dictionary learning curves with the tree greedy algorithms. 
	  The left graph corresponds to dictionary $\vDn$ and 
	  the right graph to dictionary $\vDv$.}
  \label{lcn}
\end{figure}

For this test 
we randomly take 512  beats from the $30000$ $\NN$
 heartbeats in the training set to construct a matrix $\vDn$ 
of size
 $256 \times 512$, which is used as initial dictionary. 
The learning curves for the dictionary $\vDn$ 
with the 3 greedy algorithms are shown in left graph of Fig.\ref{lcn}. 
We repeat the process but taking the 512 
dictionary atoms randomly from the 
$3358$ $\VV$ beats in the training set. The 
learning curves for this dictionary are 
  shown in the right graph of Fig.\ref{lcn}.

The classification performance is assessed by  means of the
 true positive (TP), false positive (FP), and false negative (FN)
 outcomes. These values
are used to calculate the following statistics metrics 
for each class.\\

Sensitivity ($\SE$): Number of correctly classified heartbeats
among the total number of beats in the set, i.e.

$$\SE= \frac{\TP}{\TP + \FN}.$$

Positive predictivity (PP): Ratio of correctly classified heartbeats   to all the beats classified in that class.

$$\PP= \frac{\TP}{\TP + \FP}.$$

Additionally, the total accuracy ($\AC$) of the classification is 
calculated as the fraction of correctly classified heartbeats in
both classes

$$\AC=\frac{\text{Correctly classified beats in both classes}}{\text{Total number of beats}}.$$

In Table \ref{StT} these scores are given 
 as the mean value 
of 5 realisations corresponding to 5 random initialisations 
in the dictionary learning  process. The 
standard deviations (std) are shown in rows 4,6,8,10, and 12 
of Table \ref{StT}. The Matlab codes for implementing the approach 
with the all the tree greedy algorithms have been made available 
on \cite{webpage2}.\\

\begin{table}[!ht]
\begin{center}
\begin{tabular}{|l||c|c|c|c||c|c|c|c||c|c|c|c||}
\hline

	& \multicolumn{4} {|c||}{MP}& \multicolumn{4} {c||}{OMP}& 
	\multicolumn{4} {|c||}{OOMP} \\\hline \hline
	Crit.	& I (a)& I (b) & II & III& I (a) & I (b) & II & III&
	 I (a)& I (b) & II & III\\ \hline \hline
	$\SEN$ (\%)&{\bf{99.7}}&{\bf{99.7}}&{\bf{99.7}}& {\bf{99.7}}&99.1&99.1 &99.3&{\bf{99.4}}&{\bf{98.8}}&{\bf{98.8}}&98.5& 97.4\\\hline
std &0.05&0.05&0.04&0.03&0.09&0.09&0.01&0.04&0.14&0.13&0.16&0.45\\ \hline \hline
	$\SEV$ (\%)&95.7&95.7&{\bf{96.9}} &{\bf{97.6}}&
	96.5&96.5&96.7&{\bf{97.0}}&
	{\bf{97.6}}&{\bf{97.6}}&97.3&95.3\\\hline
        std &0.65 &0.63&0.45 &0.34&
	0.24&0.18&0.27&0.25&
	0.35&0.21&0.47&1.17\\ \hline \hline
	$\PPN$ (\%)&99.7&99.7&99.8&{\bf{99.9}}& {\bf{99.8}} &{\bf{99.8}}&{\bf{99.8}}&{\bf{99.8}}&{\bf{99.9}}&{\bf{99.9}}&99.8&99.7\\ \hline 
        std &0.04&0.04&0.03&0.02&0.01&0.01 &0.02&0.01&0.02&0.01&0.03&0.07\\ \hline \hline
	$\PPV$ (\%)&{\bf{94.9}}&{\bf{94.9}}&94.6&94.1&86.4&86.6&{{88.6}}&{\bf{90.4}}&{\bf{82.5}}&{\bf{82.5}}&79.5&68.4\\\hline
        std &0.71&0.71&0.55&0.48&1.27&1.27&0.17&0.55&1.56&1.72&1.72&3.48\\ \hline \hline
	$\AC$ (\%)&{\bf{99.5}}&{\bf{99.5}}&{\bf{99.5}}&{\bf{99.5}}&99.0&{99.0}&99.1&{\bf{99.3}}&{\bf{98.7}}&{\bf{98.7}}&98.5&97.3\\\hline
        std &0.03&0.03&0.02&0.03&0.01&0.01&0.02&0.05&0.12&0.13&0.12&0.38\\ \hline \hline
\end{tabular}
	\caption{Average statistics scores (of 5 different initialisations in 
	the dictionary learning procedure) 
	 for classes $\NN$ and $\VV$, obtained with MP, OMP, and OOMP,  and the  sparsity criteria discussed in Sec.~\ref{SM}. std indicates the corresponding standard deviations and  $\AC$ the average accuracy.} 
	 \label{StT}
	\end{center}
\end{table}

\newpage
{\bf{Discussion of results}}\\
\noi

\noi
i) The high accuracy and $\SE$ values for both classes, 
indicates that:  
35\% of the whole $\NN$  beats in the data set and 
50\% of the whole $\VV$ beats in the data set provide enough 
examples in the training set to learn dedicated dictionaries for each class. \\

\noi
ii) From Table I we can assert that each of the greedy algorithms 
performs better with a particular decision criterion.\\ 

\noi
iii) MP, combined with the decision Criterion II (norm-1), produces 
the highest statistics scores. \\

\noi
iv) For most of the scores, the random initialisation of the learning process does not change the results in any significant manner (low std values). \\

\noi
v) The comparatively much lower values  of $\PPV$  
 reflex the enormous difference of the 
number of $\NN$  and $\VV$ beats in the testing sets 
(57277 vs  3359  beats). Hence, even if the percentage of incorrectly classified $\NN$ beats is small, the classification being 
binary implies that the number of incorrectly classified $\NN$ beats count as false positive $\VV$. This causes the low $\PPV$ values in comparison to the other  statistics metrics.
\\ 

\noi
vi) Even if the OOMP approach secures the least number of atoms in th learning process (c.f. Fig.\ref{lcn}) it is not the approach rendering the highest classification scores. This is because  what matters for classification is the relative sparsity with respect to the 
2 dictionaries. It is not surprising, though, that OOMP is the only approach that works better with Criterion I. 

\begin{remark}
Accuracy of 99\% for differentiating classes in the whole 
	MIT-BIH Arrhythmia  data set is state of the art result obtainable from 
	other features and  other 
	 machine learning classifiers. See for instance\\ 
	 \begin{itemize}
	 \item[\cite{MAP13}] $\AC\,\, 99\%$for classification 
			 of 2 classes,
	 \item[\cite{AS15}] $\AC \,\,99\%$  for classification  of 5 classes,
	 \item[\cite{RRS16}] $\AC\,\, 99.2\%$ for classification  of 16 classes,
	 \item[\cite{RR18}] $\AC\,\, 99.3\%$ for classification  of 16 classes. 
	 \end{itemize}
Although results are not strictly comparable, because the number of 
classes  are different, we believe it appropriate to highlight 
	that the 
99.5\% accuracy attained in this work achieves equivalent   
	values as other approaches do for binary and
	multi classification.
\end{remark}
\noi
\subsection{Test II}
We test now the proposal in a more challenging situation. 
Because the morphology of $\NN$ and $\VV$ depends on the 
particular ECG signal, a realistic test is carried out
 by taking the training set and testing set from different records (corresponding to different patients). 

Following previous publications \cite{COR04, RR18}
the data are divided for training and testing as below.

The training set is taken only from the 21 records below 
in the the MIT-BIH Arrhythmia  data set.

\noi
101, 106,  108, 112, 114, 115, 118, 119, 122, 124, 201, 203, 205, 207, 208, 209, 215, 220, 223, 230.\\

The remaining 21 records provide the testing set. These are 
the records\\

\noi
100, 103, 105, 111, 113, 117, 121, 123, 200, 202, 210, 212, 213, 214, 219, 221, 222, 231, 232, 233, 234
in the  MIT-BIH Arrhythmia  data set.\\

\begin{table}[!ht]
\begin{center}
\begin{tabular}{|l||c|c|c|c||c|c|c|c||c|c|c|c||}
\hline
        & \multicolumn{4} {|c||}{MP}& \multicolumn{4} {c||}{OMP}&
        \multicolumn{4} {|c||}{OOMP} \\\hline \hline
        Crit.   & I (a)& I (b) & II & III& I (a) & I (b) & II & III&
         I (a)& I (b) & II & III\\ \hline \hline
	$\SEN$ (\%)&89.8 & 89.9  & 90.5 & {\bf{91.8}}
	&88.8 & 89.5  & 88.8 & {\bf{92.5}}
	&88.7 & {\bf{89.0}}  & 85.7 & 81.5\\ \hline
std	&1.43 &  1.39 & 1.38 &1.26
	&1.39 &  1.28 & 1.48 &1.19
	&0.92 &  0.87 & 0.98 &2.10 \\ \hline \hline
$\SEV$ (\%) &87.6 & 87.6 & 90.2 & {\bf{91.0}}
	&89.8  &89.5 & {\bf{90.5}} & 88.9
	&92.3 & {\bf{92.6}} & 89.6 & 84.7 \\ \hline 
std	&1.57& 1.64& 0.75& 1.14
	&1.40& 1.24& 1.46& 1.14
	&0.76& 0.94& 1.15& 2.83 \\ \hline \hline
	$\PPN$ (\%)&99.0& 99.0& 99.2& {\bf{99.3}}
	&{\bf{99.2}}&{\bf{99.2}}& {\bf{99.2}}& 99.1
	&{\bf{99.4}}& {\bf{99.4}}& 99.1& 98.7 \\ \hline 
std	&0.11& 0.12& 0.05& 0.08
	&0.11& 0.09& 0.11& 0.08
	&0.06& 0.07& 0.10& 0.22 \\ \hline \hline
$\PPV$	&38.8 & 39.1& 41.1 &{\bf{45.1}}
	&37.1 & 38.5& 37.3 &{\bf{46.6}}
	&37.5 & {\bf{38.2}}& 31.5 &25.2 \\ \hline 
std	&3.01& 2.96 &3.30 &3.56
	&2.70& 2.70 &2.73 &4.10
	&1.96& 1.85 &1.47 &1.69 \\ \hline \hline
$\AC$	&89.7& 89.8 &90.4 &{\bf{91.8}}
	&88.9& 89.5 &88.9 &{\bf{92.2}}
	&89.0& {\bf{89.3}} &86.0 &81.7 \\ \hline 
std	&1.25& 1.21 &1.24 &1.10
	&1.24& 1.14 &1.33 &1.08
	&0.84& 0.79 &0.91 &1.81 \\ \hline \hline
\end{tabular}
\caption{Same description as in Table~\ref{StT} but for the 
	numerical Test II, in which 
training and testing sets correspond to different 
	records.}
\label{StT2}
\end{center}
\end{table}

\noi
{\bf{Discussion of results}}\\

\noi
i) As expected, the classifications scores are lower than in 
Test I. This is because not all the shapes of the beats in the 
testing set  
bears similarity with shapes present in the training set. \\

\noi
ii) Also in this test each of the greedy algorithms
performs better with a particular decision criterion.\\

\noi
iii) The greedy algorithm OOMP is the only one that performs 
better with Criterion I.\\

\noi
iv) Clearly MP works best with Criterion III  and for  most scores 
OMP as well. \\

\noi
v) The low values of $\PPV$ have the same cause as in Test I. 
Here all values are even lower because the incorrectly classified 
$\NN$ beats are more than in Test I.\\

\begin{remark}
In order to put the results into context we give here some 
	indication of 
	other technique performance. Nonetheless, it should be stressed once again that scores are not strictly comparable due to 
	difference in the number of classes and, for the binary case, 
	the classes and records for learning and testing.
\begin{itemize}	
	\item[\cite{AY19}] $\AC \,\, 97\%$ for binary classification with 37 records for 
		training and 11 records for testing, two of which are from a different data set.
	\item[\cite{ZLC14}] $\AC  \,\,86.7\%$ for classification of 4 classes.
	\item[\cite{RRS16}] $\AC  \,\,89.18 \%$ for classification of 5 classes.
	\item[\cite{RR18}]  $\AC  \,\,89.93 \%$ classification in of 5 classes.
\end{itemize}
\end{remark}
\section{Conclusions}
\label{conclu}
A set up for binary morphological identification of 
heartbeats on the basis of sparsity metrics has been laid out. 
The proposal was tested for identification of $\NN$ and $\VV$ 
beats in the MIT-BIH Arrhythmia data set, achieving 
state of the art scores. 

Because the number of $\NN$ and $\VV$ beats in the
 MIT-BIH Arrhythmia are unbalanced,
 the $\AC$ score is not reliable for assessment. Nevertheless,
the percentage of correctly classified $\NN$ and $\VV$ are 
similar to the $\AC$ score
 in both numerical tests.

The results are encouraging because the proposed binary 
identification is realised outside the usual machine leaning 
framework, using sparsity as a 
{\em{single}} parameter for making a decision. Thus, extensions of 
the approach to allow for combination with other features and
other  machine leaning techniques are readily foreseen. 

 The possibility of implementing the technique depends on the 
 availability of `enough' examples  to learn the 
 corresponding dictionaries. In the numerical tests 
 realised here, for instance, 3359 examples for beats $\VV$ were 
 enough to learn the dictionary $\vDv \in \R^{256 \times 512}$.

\end{document}